\documentclass[aps,prl,twocolumn,showpacs,floatfix,superscriptaddress]{revtex4-1}

\usepackage[dvips]{graphics}
\usepackage{epsfig}
\usepackage{times}
\usepackage{amsmath,amssymb}
\usepackage{amsthm}
\usepackage{subfigure,overpic}
\usepackage{mathrsfs}
\usepackage{multirow}

\def \beq {\begin{equation}}
\def \edq {\end{equation}}
\def \bes {\begin{subequations}}
\def \eds {\end{subequations}}
\def \beqn {\begin{equation*}}
\def \edqn {\end{equation*}}

\def \up {\uparrow}
\def \down {\downarrow}

\def \veps {\varepsilon}

\begin{document}
\title{Fluctuation Relations for Spintronics}
\author{Rosa L\'opez}
\affiliation{Departament de F\'{i}sica, Universitat de les Illes Balears,
  E-07122 Palma de Mallorca, Spain}
\affiliation{Institut de F\'{i}sica Interdisciplinar i de Sistemes Complexos
  IFISC (CSIC-UIB), E-07122 Palma de Mallorca, Spain}
\author{Jong Soo Lim}
\affiliation{Institut de F\'{i}sica Interdisciplinar i de Sistemes Complexos
  IFISC (CSIC-UIB), E-07122 Palma de Mallorca, Spain}
\author{David S\' anchez}
\affiliation{Departament de F\'{i}sica, Universitat de les Illes Balears,
  E-07122 Palma de Mallorca, Spain}
  \affiliation{Institut de F\'{i}sica Interdisciplinar i de Sistemes Complexos
  IFISC (CSIC-UIB), E-07122 Palma de Mallorca, Spain}

\begin{abstract}
  Fluctuation relations are derived  in systems where the spin degree of freedom and magnetic interactions play a crucial role.  
  The form of the non-equilibrium fluctuation theorems relies in the assumption of  a local balance condition. We demonstrate that 
  in some cases the presence of magnetic interactions violates this condition. Nevertheless, fluctuation relations can be obtained 
  from the micro-reversibility principle sustained only at equilibrium as a symmetry of the cumulant generating function 
  for spin currents. We illustrate the spintronic fluctuation relations for a quantum dot coupled to  
  partially polarized helical edges states .
\end{abstract}
\maketitle
\paragraph{Introduction.} Non-equilibrium fluctuation theorems (FTs) 
\cite{PhysRevLett.78.2690,PhysRevE.60.2721,RevModPhys.81.1665}, widely 
used for macroscopic systems, are based on the thermodynamics governing the physical processes when they evolve forward and 
backward in time.  The boundary conditions for the forward and the time-reversed  processes determine the balance condition for the 
entropy exchange and therefore the form of the fluctuation theorem \cite{RevModPhys.81.1665}.  
 The applicability of the non-equilibrium FTs to quantum systems has become an exciting problem and, in  
 particular, to the case of the charge transfer phenomena in mesoscopic systems 
 in the context of the full counting statistics \cite{levitov,PhysRevLett.88.196801,PhysRevLett.90.206801,1742-5468-2006-01-P01011}.
 Interestingly, relations akin to the  fluctuation-dissipation 
theorem \cite{PhysRev.32.110,PhysRev.32.97,JPSJ.12.570,Einstein.1905}
have been formulated beyond the linear response regime 
\cite{PhysRevB.72.235328,
1742-5468-2006-01-P01011,PhysRevLett.101.046802,PhysRevB.78.115429,PhysRevLett.101.136805,%
san09,san10,lim10,gol11,bul11,kra11,gan11}.
These \textit{fluctuation relations} relate nonequilibrium fluctuation and dissipation
coefficients for phase-coherent conductors.
However, the role of a genuine quantum property such as the \textit{spin} degree of freedom
in the fluctuation relations has not been yet investigated in detail. Our motivation is not only fundamental
since the electronic spin offers enormous advantages to 
create devices with unusual and extraordinary new functionalities \cite{fab07}.  
The purpose of this work is thus to generalize the \textit{fluctuation relations}
for \textit{spintronic} systems.
\begin{figure}[!t]
  \centering
 \includegraphics[width=0.4\textwidth, angle=90,clip]{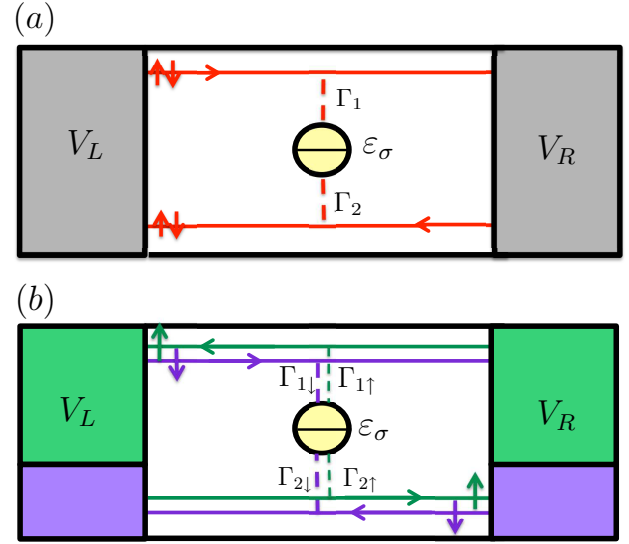}
  \caption{(Color online) (a) Sketch of a quasi-localized level ($\veps_\sigma$) coupled to chiral edge states with $\Gamma_{1,2}$ and driven out of equilibrium with $V_L$ and $V_R$ bias voltages.  For $B>0$ the upper (lower) edge state is injected from $V_L$ ($V_R$).  (b) Localized level coupled to unequally spin populated helical edge states due to the spin injection from the ferromagnetic leads. Then, the tunnelling couplings are spin-dependent: $\Gamma_{(1,2)\uparrow(\downarrow)}$.  Ferromagnetic electrodes: larger light area (green)  corresponds to majority spins whereas smaller dark area (purple) are minority spins. Helical states: upper left (upper right) movers are spin up (spin down) carriers injected from $V_R$ ($V_L$).}
  \label{fig:1}
\end{figure}

Fluctuation relations are generated from the cumulant generating function (CGF) $\mathcal{F}(\chi)=\ln \sum_Q P(Q,t)e^{-i\chi Q}$ where $P(Q)$ is the charge distribution function. Firstly, the CGF $\mathcal{F}$ is expanded in a Taylor expansion in terms of affinities $A=\left(qV_1/k_BT, qV_2/k_BT,\cdots\right)$ ($q$ is the electron charge, $k_B$ is the Boltzmann constant, $T$ is the temperature and $V_{i}$, $i=1,2\dots$ are the applied voltages) and counting fields $\chi=\left(\chi_1,\chi_2,\cdots\right)$ around the equilibrium condition. Then, thanks to the symmetries $F(0,A) = 0$
(probability conservation condition) and $F(-A,A) = 0$ (global detailed balance condition) fluctuation relations among the higher-order non-linear cumulants are found. Indeed, the symmetries of $\mathcal{F}$ can be considered as the non-equilibrium FT versions for the currents within a transport theory. Initial
experiments by using a mesoscopic dot interferometer have tested these relations \cite{PhysRevLett.104.080602}. In this experiment, the noise susceptibility and the second order conductance were found to be proportionally related. 

Spins are sensitive to magnetic fields and also to electric fields due to spin-orbit interactions. Fluctuation relations 
for the charge transport have been  formulated in the presence of magnetic fields \cite{PhysRevB.78.115429,
PhysRevLett.101.136805,san09}. In Ref. \cite{PhysRevB.78.115429} the non-equilibrium FT for the forward and backward charge distribution probability at opposite $B$ polarities $P(Q,B)/P(-Q,-B)=e^{Q A}$  was used to derive such fluctuation relations. However, some caution is needed since 
$P(Q,B)$ and $P(-Q,-B)$ are considered for a system driven
out of equilibrium in which the interacting internal potentials are no longer even functions of $B$ \cite{PhysRevLett.93.106802} and 
the application of such theorem may break down \cite{PhysRevLett.101.136805}. To circumvent this obstacle,  
Ref. \cite{PhysRevLett.101.136805} uses a symmetry of $\mathcal{F}$ associated with the micro-reversibility condition \textit{only} 
at equilibrium $P(Q,B)_{A\to 0}=P(-Q,-B)_{A\to 0}$. We here derive the \textit{spintronic fluctuation relations}  in the same spirit 
when time-reversal symmetry is broken not only by external magnetic fields but also by the presence of 
ferromagnetic electrodes. In this case, at equilibrium $P(Q,B,p)_{A\to 0}=P(Q,-B,-p)_{A\to  0}$, where $p$ is the 
lead magnetization \cite{ada06}. 

We illustrate our findings with a quasi-localized level coupled to helical edge states which are partially polarized by the presence of polarized electrodes [see Fig. 1 (b)]. Helical modes have been observed in topological insulators \cite{RevModPhys.82.3045} and proposed to occur in quantum wires \cite{PhysRevB.84.195314} and in carbon nanotubes \cite{PhysRevLett.106.156809}. This quantum spin Hall state consists of gapless excitations that exist at the boundaries in which its propagation direction is correlated with its spin due to the spin-orbit interaction. By electrostatic gating, quasi-localized states can form in the interior of the carbon nanotubes and quantum wires.  Furthermore, ferromagnetic contacts have been successfully  attached to these nanodevices \cite{ferroleads}.  Finally, Ref.\ \cite{spinhallQD} suggests to create a quasi-bound state in quantum spin Hall setups by using ferromagnetic insulators that serve as tunnelling barriers.


\paragraph{Local Detailed Balance.}
Consider a system described by a set of  $m$ discrete states coupled to $\ell$-electronic reservoirs. We assume that its 
dynamics is governed by the master equation $d\rho/dt=\mathcal{W} \rho$, where $\mathcal{W}$ is  the transition rate matrix, and $\rho$ 
denotes the occupation probabilities for the $m$-states.  The exchange of energy ($\Delta E^{(\ell)}$) or particles ($\Delta 
N^{(\ell)}$) in the $\ell$-th-reservoir with inverse temperature $\beta^{(\ell)}$
is described by adding counting fields ($\chi_E^{(\ell)}$, $\chi_N^{(\ell)}$) to the 
off-diagonal matrix elements of $\mathcal{W}$. Thus, for the upper off-diagonal the transition rate from the state $n$ to the state $m$ are modified according to $W_{nm}=\sum_{\ell} W^{(\ell)}_{nm} e^{\chi_E^{(\ell)}\Delta E^{(\ell)}+\chi_N^{(\ell)}\Delta N^{(\ell)}}$  (for $n<m$)  whereas for the lower 
off-diagonal terms these rates are $W_{nm}=\sum_{\ell} W^{(\ell)}_{nm} e^{-\chi_E^{(\ell)}\Delta E^{(\ell)}-\chi_N^{(\ell)}\Delta N^{(\ell)}}$ ($n>m$).  
Usually,  boundary conditions are taken into account through the \emph{local detailed balance} (LDB) condition in which weight factors 
$e^{-\beta^{(\ell)} (\mathcal{H_\ell}-\mu N_\ell)}$  ($\mathcal{H_\ell}$, and $N_\ell$ denote the Hamiltonian and the particle number operator, 
respectively for the $\ell$th-reservoir) balance forward and backward processes. To be more specific:
\beq
\label{ldb}
\frac{W^{(\ell)}_{nm}}{W^{(\ell)}_{mn}}=e^{-\beta^{(\ell)}(\Delta E^{(\ell)}-\mu^{(\ell)}\Delta N^{(\ell)})}\,.
\edq
From the LDB condition the equality $\mathcal{W}(\chi_E^{(\ell)}, \chi_N^{(\ell)})=\mathcal{W}^T(\beta^{(\ell)}-\chi_E^{(\ell)}, 
-\beta^{(\ell)}\mu^{(\ell)}-\chi_N^{(\ell)})$ is automatically satisfied reflecting the following symmetry for the generating function 
$\mathcal{F}$ [which is constructed from $\mathcal{W}(\chi_E^{(\ell)},\chi_N^{(\ell)}$)]:
\beq\label{symmetry}
\mathcal{F}[\chi_E^{(\ell)}, \chi_N^{(\ell)})] =\mathcal{F}[\beta^{(\ell)}-\chi_E^{(\ell)}, -\beta^{(\ell)}\mu^{(\ell)}-\chi_N^{(\ell)}]\,.
\edq
Although in many systems we can assume some type of LDB condition, in general,  Eq.(\ref{ldb}) is not fulfilled \cite{vai2007}. To see this in a quantum conductor,  we consider the system sketched in Fig. 1(a) in which the presence of a magnetic field breaks time-reversal symmetry. The system consists of a quasi-localized state with energy $\veps_d$ in the Coulomb blockade regime coupled to two chiral states propagating along the opposite edges of a quantum Hall conductor  (filling factor $\nu = 1$) \cite{PhysRevLett.93.106802,san09,wang2011}. In the infinite charging energy limit case only two dot charge states are permitted: $|0\rangle $ and $|1\rangle$. For positive magnetic fields $B>0$ carriers in the upper (lower) edge state move from the left (right) terminal to the right (left) terminal. The current flow is reversed for $B<0$. Interaction between the quasi-localized state and the edge states takes place via tunnel couplings $\Gamma_1$ and $\Gamma_2$ and capacitive couplings $C_1$ and $C_2$. The chiral coupling involves different transition rates  depending on the polarity of the magnetic field. For a positive $B$ we have  $W_{0 1}^{L(R)}=\Gamma_{1(2)} f(B,\mu_{L(R)})$,  
$W_{10}^{L(R)}=\Gamma_{1(2)}[1-f(B,\mu_{L(R)})]$,  where 
$f(B,\mu_{L(R)})=1/(1+\exp{\beta[\mu_d(B)-\mu_{L(R)}]})$ is the Fermi-Dirac distribution function, $\mu_{L(R)}=qV_{L(R)}+E_F$ denotes the electrochemical potential  in the lead $L(R)$ with $E_F=0$ the Fermi energy, and $\mu_d$ is the electrochemical 
potential of the quasi-localized state which is self-consistently calculated and depends on the $B$ 
orientation \cite{san09}. For $B > 0$, $\mu_{d}(B) -\mu_L= \veps_{d}-(1-\eta)qV/2$, where $\eta=(C_1-C_2)/(C_1+C_2)$ is the 
capacitance asymmetry parameter and $V=V_L-V_R$. For $B<0$ the motion of the edge states is reversed and then $\mu_{d}(-B)-\mu_L = 
\veps_{d} -(1+\eta)qV/2$. Because of the fact that $\mu_{d}(B)\neq \mu_{d}(-B)$, the LDB condition is not satisfied. Clearly,
\beq
\frac{W^L_{10}(B)}{W^L_{01}(-B)}=e^{\beta (\veps_d- q V/2)}[1-\beta\eta \frac{qV}{2} \tilde{f}_{eq}+\mathcal{O}(V^2)]\,, 
\edq
where $\beta$ is the common inverse temperature and $\tilde{f}_{eq}=1-2f_{eq}$ with $f_{eq}$ being the Fermi function at equilibrium ($V_L=V_R$) . 
Importantly,  the violation of the LDB condition occurs for asymmetric capacitance couplings only. In the symmetric case or at equilibrium, Eq. (1) is recovered. The violations of LDB are thus a consequence of \textit{asymmetric, chiral states out of equilirium}
\begin{figure}[!t]
  \centering
 \includegraphics[width=0.25\textwidth, angle=90]{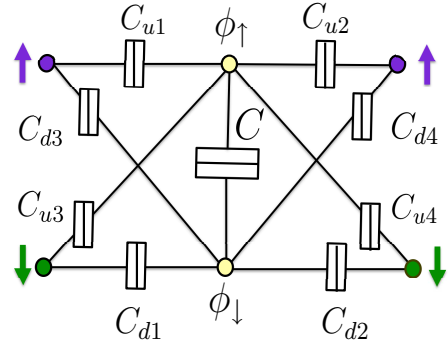}
  \caption{(Color online) Electrostatic model for the quasi-localized state connected to helical spin edge states. Spin up(down)  localized level is capacitatively coupled to the upper up(down) helical channel with capacitances $C_{u1(u3)}[C_{d1(d3)}]$ and  to the lower up(down) helical channel with capacitances $C_{u2(u4)}[C_{d2(d4)}]$. A mutual capacitance between up and down localized levels is accounted for with $C$. $\phi_{\up(\down)}$ denotes the spin up (down)  internal potential for the quasi-bound state.}
  \label{fig:2}
\end{figure}

We now show that violations of the LDB conditions are also present in the absence of magnetic fields and when the spin degree of freedom is explicitly accounted for. For that purpose we consider the system sketched in Fig. 1(b), a quasi-bound state which is tunnel coupled to helical edge states. The helical modes are partially polarized due to their coupling to two
ferromagnetic electrodes with parallel magnetization and equal
polarization $p$. In this manner, polarized helical edge states
are described with a spin-dependent density of states (DOS) 
$D_{i s}=(1+s p)D_i/2$, where $s=+(-)$ for
$\uparrow(\downarrow)$-helical mode
and, $D_i$, with $i=u,d$ denoting the upper and lower edge-state DOS 
in the absence of polarization \cite{PhysRevB.53.2064}. In the wide band limit
approximation, the tunnelling rates become spin-dependent,
$\Gamma_{i s}=\pi |t_i|^2 D_{i s}$, with $|t_i|^2$ the tunnel
probability from the $i$-th edge state. Defining
$\Gamma_i=\pi |t_i|^2 D_i$, we find
$\Gamma_{i s}=(1+s p) \Gamma_i /2$. 

Our transport description also includes an electrostatic model for interactions
between the dot and the edge states. Within the mean-field approach,
the electrochemical capacitive coupling $C_\mu$ consists of a
geometrical capacitance contribution, $C_g$, which depends on the
width and the height of the tunnel barrier, and a quantum capacitance term, $C_{is,q}$,
which we take as proportional to $D_{is}$:
\beq\label{electrochemical}
C_{\mu i s}^{-1}=\frac{1}{C_g}+\frac{1}{q^2 D_{is}}\,.
\edq
We emphasize that the capacitive couplings are, in general,
spin dependent~\cite{mac99}. For sufficiently large geometrical
capacitances, $C_g\gg C_{is,q}$, we find from Eq. (4) the capacitative couplings 
\beq
C_{u1(2)}=\frac{1+p}{2} C_{1(2)}, \,\, C_{u3(4)}=\frac{1-p}{2} C_{3(4)}\,,
\edq
where $C_{i}=q^2 D_i$ with $C_{u1(d1)}$, $C_{u2(d2)}$ being the capacitive couplings
between left (right) movers with up (down) spin along the top edge
and the dot electron with spin $\uparrow$ ($\downarrow$) whereas
$C_{u3(d3)}$ and $C_{u4(d4)}$ couple the same dot state
with right (left) movers along the bottom edge with up (down) spins
(see Fig. 2).
For thin edge states, $D_i$ depends on the steep confinement
potential at the top and bottom edges, which will generally differ \cite{PhysRevB.53.2064}.
Then, we take $C_1=C_3$ and $C_2=C_4$ but $C_1\neq C_2$.
As a consequence, the capacitive couplings between the dot
and the edges is asymmetric: $\eta\neq 0$. Furthermore, since the upper
and lower edge modes are equally polarized, one has $C_{dj}=C_{uj}$.

Consider for the moment the case where the capacitance coupling
between the dot states is neglected ($C=0$). Then, we calculate
the {\em spin-dependent} electrochemical potential of the dot
and find the simple relation $\mu_\sigma (p)-\mu_{\bar\sigma} (-p)=p\eta
V$. Now, in a time-reversal operation we have to invert the lead polarization,
the edge state spin index, and
the dot spin. Doing so, we obtain an invariant result only at equilibrium
($V=0$) or for symmetric capacitive couplings. But, in general, when $V\neq 0$
the original state is not restored and, as a consequence,
LDB is not fulfilled:
\begin{eqnarray}
\frac{W_{0\sigma}(p)}{W_{\bar\sigma 0}(-p)}=
e^{\beta[\veps_d- qV/2]}[1- \beta\eta p \frac{q V}{2} \tilde{f}_{eq}+\mathcal{O}(V^2)],
\end{eqnarray}
where $\sigma=\{\uparrow,\downarrow\}$ denotes the dot spin index.
We stress that helicity is needed in our example to find departures from LDB. Although we cannot
rule out the possibility that nonchiral, spintronic systems (e.g., a dot directly attached
to ferromagnetic leads) might show such departures if coherent tunneling or strong
correlations are taken into accoutn, our conceptually simple system already exhibits the effect
with fully analytical expressions.

\paragraph{Spintronic fluctuation relations.}
We now treat on equal footing the  presence of both magnetic fields and polarized contacts.  The spin-dependent probability distribution satisfies the micro-reversibility condition, but only at equilibrium
\begin{multline}\label{micro-reversibility}
P(\{n_{\alpha{s}},n_{\beta{s'}},\ldots\};B, p)\\
 = P(\{-n_{\alpha{\bar{s}}},-n_{\beta{\bar{s'}}},\ldots\};-B, -p)\,,
\end{multline}
where $\alpha$ and $s$ are the lead and spin indices, respectively and $p\equiv (p,p',\cdots)$ contains the magnetizations for the 
leads. The CGF $\mathcal{F}(\{ i\chi \}, A)$ can be expanded in terms of powers of voltages and counting fields 
\beq
\mathcal{F}(\{i\chi\},A) = \sum_{\{k_{\alpha s}\},\{l_{\alpha }\}} f_{\{k_{\alpha s}\},\{l_{\alpha }\}} \frac{\prod_{\alpha s} (i\chi_{\alpha s})^{k_{\alpha s}}\prod_{\alpha} A_{\alpha}^{l_{\alpha}}}{\prod_{\alpha} k_{\alpha s}! \prod_{\alpha } l_{\alpha }!}
\label{eq:Fexp}
\edq
and
\beq
f_{\{k_{\alpha s}\},\{l_{\alpha }\}} = \prod_{\alpha s} \partial^{k_{\alpha s} + l_{\alpha }} \mathcal{F}(\{i\chi\}, A) / \partial(i\chi_{\alpha s})^{k_{\alpha s}} \partial A_{\alpha}^{l_{\alpha}} \Big|_{i\chi\to 0,A \to 0}
\edq
where $k$ and $l$ are non-negative integers.  From the derivatives of $\mathcal{F}(\{i\chi\}, A) $ with respect to the counting fields, the cumulants are generated. In this way, the average current through terminal $\alpha$ with spin $s$ is derived from Eq. (\ref{eq:Fexp}) as
$\langle I_{\alpha s}\rangle =f_{\{1_{\alpha s}\}}$. Similarly, second cumulant (current-current correlation) $S_{\alpha 
s,\beta s^\prime}\equiv \langle \Delta I_{\alpha s} \Delta I_{\beta s^\prime}\rangle$ ($\Delta I_{\alpha s}=\hat 
I_{\alpha s}-\langle I_{\alpha s}\rangle)$, where $\hat{I}$ denotes the current operator)  and the third cummulant 
$C_{\alpha s\beta s^\prime \gamma s^{\prime\prime}}\equiv \langle \Delta I_{\alpha s} \Delta I_{\beta s^\prime}\Delta 
I_{\gamma s^{\prime\prime}}\rangle$ are given by  $S_{\alpha s\beta s^\prime}=f_{\{1_{\alpha s} 1_{\beta 
s^\prime}\}}$ and $C_{\alpha s\beta s^\prime \gamma s^{\prime\prime}}=f_{\{1_{\alpha s} 1_{\beta s^\prime} 
1_{\gamma s^{\prime\prime}}\}}$, respectively. We  expand, both, the current $\langle I_{\alpha s}\rangle$, 
and  the noise $\langle S_{\alpha s\beta s^\prime}\rangle$ in powers of the applied voltages as follows
\begin{eqnarray}\label{exp12}
&&\langle I_{\alpha s}\rangle =\sum_{j} G_{\alpha s,j}^{(1)} V_j +\frac{1}{2}\sum_{j,k} G_{\alpha s,jk}^{(2)} V_j V_k +\mathcal{O}(V^3)\,, \nonumber\\ 
&&\langle S_{\alpha s\beta s^\prime}\rangle =S_{\alpha s\beta s^\prime}^{(0)} +  \sum_{j} S_{\alpha s\beta s^\prime,j}^{(1)} V_j +\mathcal{O}(V^2)\,.
\end{eqnarray}
Here $G_{\alpha s,j}^{(1)}=f_{\{1_{\alpha s}\}, \{1_{ j}\}}$ corresponds to the linear conductance, $G_{\alpha s,j k}^{(2)}=f_{\{1_{\alpha s}\}, \{1_{ j} 1_{k}\}}$ is the second-order conductance, and $S_{\alpha s,\beta s^\prime,j}^{(1)}=f_{\{1_{\alpha s} 1_{\beta s^\prime}\},\{1_j}\}$ is the noise susceptibility.
Fluctuation relations are expressions that relate the $f$-coefficients at different order in voltage. To derive explicitly these relations we employ the micro-reversibility condition at equilibrium $F(i\chi_{\alpha s},i\chi_{\beta s^\prime},\cdots ,A,+B)|_{A=0} =F(-i\chi_{\alpha \bar{s}},-i\chi_{\beta \bar{s'}},\cdots ,A,-B)|_{A=0} $ [cf. Eq.\ (\ref{micro-reversibility})]. It is convenient to define the symmetrized ($+$) and anti-symmetrized ($-$) combination of the $f$-factors
\beq\label{ref}
f_{\{k_{\alpha s}\},\{l_{j}\}}^{\pm} =f_{\{k_{\alpha s}\}, \{l_{j}\}}(B, p)\pm
f_{\{k_{\alpha \bar{s}}\}, \{l_{j}\}}(-B, -p)\,.
\edq
where $f_{\{k_{\alpha \bar{s}}\}, \{l_{j}\}}(-B, -p)$ is generated by means of time reversal operation $B\to -B$, $p\to -p$, and $s\to -s$. According to Eq.\ (\ref{micro-reversibility}) the $f^{\pm}$-factors are even(odd) functions  under time-reversal operation.  This even-odd property is translated into the following relations for the equilibrium coefficients [in the sense of a voltage expansion,  see Eqs. (\ref{exp12})]:
\begin{eqnarray}
&&S_{\alpha s \beta s^\prime}^{(0)}(B,p) = S_{\alpha \bar{s} \beta \bar{s}^\prime}^{(0)}(-B,-p)\,, \\ \nonumber
&&C_{\alpha s \beta s^\prime \gamma s^{\prime\prime}}^{(0)}(B,p) = -C_{\alpha \bar{s} \beta \bar{s}^\prime \gamma \bar{s}^{\prime\prime}}^{(0)}(-B,-p)\,.
\end{eqnarray}
Now by using the global detailed balance condition $\mathcal{F}(-A,A)_{\pm}=0$, and the probability conservation  $\mathcal{F}(0,A)_{\pm}=0$ one can derive the \textit{spintronic fluctuation relations} among different $f_{\pm}$-factors. Here we explicitly show those that relate the coefficients appearing in the third cumulant, noise and the conductances in the voltage expansion of Eq. \eqref{exp12}: 
\begin{eqnarray}\label{spintronicrelations}
&&C_{\alpha s \beta s^\prime \gamma s^{\prime\prime} \pm}^{(0)} \\
\nonumber
&=& k_B T \Big[S^{(1)}_{\alpha s\beta s^\prime,\gamma \pm} + S^{(1)}_{\alpha s\gamma s^{\prime\prime}, \beta \pm}
+S^{(1)}_{\beta s^\prime \gamma s^{\prime\prime}, \alpha \pm} \nonumber \\
&-& k_B T \left (G^{(2)}_{\alpha s,\beta \gamma \pm} +
G^{(2)}_{\beta s^\prime,\alpha \gamma \pm}+G^{(2)}_{\gamma s^{\prime\prime}, \alpha \beta \pm} \right)\Big]\,. \nonumber
\end{eqnarray}
Fluctuation relations between even higher-order response coefficients toward the strongly
nonequilibrium domain can be similarly found, relating different current cumulants
at different order; however, the resulting expressions, already in the spinless case,
look rather cumbersome \cite{PhysRevLett.101.136805}.

We verify Eq. \eqref{spintronicrelations} in a multi-terminal setup in which LDB condition is broken. 
For that purpose we generalize the two terminal quantum spin Hall bar system [Fig. 1(b)] 
to the multi-terminal case in which upper and lower helical modes are now connected to 
different terminals $V_{i}$, $i=1\cdots 4$ [see inset in Fig. 3(d)]. We additionally consider spin-flip 
relaxation events within the quasi-bound state  that can occur due to spin-spin interactions 
with a spin fluctuating environment  (hyperfine interaction, spin-orbit interactions, etc.). 
We phenomenologically model this rate as $\gamma_{sf}^{\sigma\bar\sigma} = \gamma_{sf} 
\exp\left[(\veps_{\sigma}-\veps_{\bar\sigma})/(2k_B T)\right]$.  Notice that due to spin-flip 
events spin up and down currents are correlated and then Eq. (\ref{spintronicrelations}) is 
satisfied in a non-trivial manner. We emphasize that Eq. (\ref{spintronicrelations}) 
is verified (see Fig. 3) even for a finite capacitance asymmetry where the LDB condition is not met.
\paragraph{Conclusions.}
In short, we have shown that the applicability of non-equilibrium  FT when magnetic interactions are present is not \textit{a priori } ensured. We  illustrate this statement by using a quasi-localized level coupled to a chiral one-dimensional conducting channels. We demonstrate that local detailed balance condition is not satisfied when a magnetic field is included and the system is driven out of equilibrium. Importantly, we have derived the \textit{fluctuation relations for spintronic systems} and have explicitly verified them in the illustrative case of a quasi-localized state coupled to partially polarized helical edge states.
Our formalism is based on zero-frequency fluctuations and time-independent fields
but in the presence of arbitrary interactions. Promising
avenues for future work include finite-frequency calculations and ac fields.
\begin{figure}
\centering
\includegraphics[width=0.4\textwidth, angle=90]{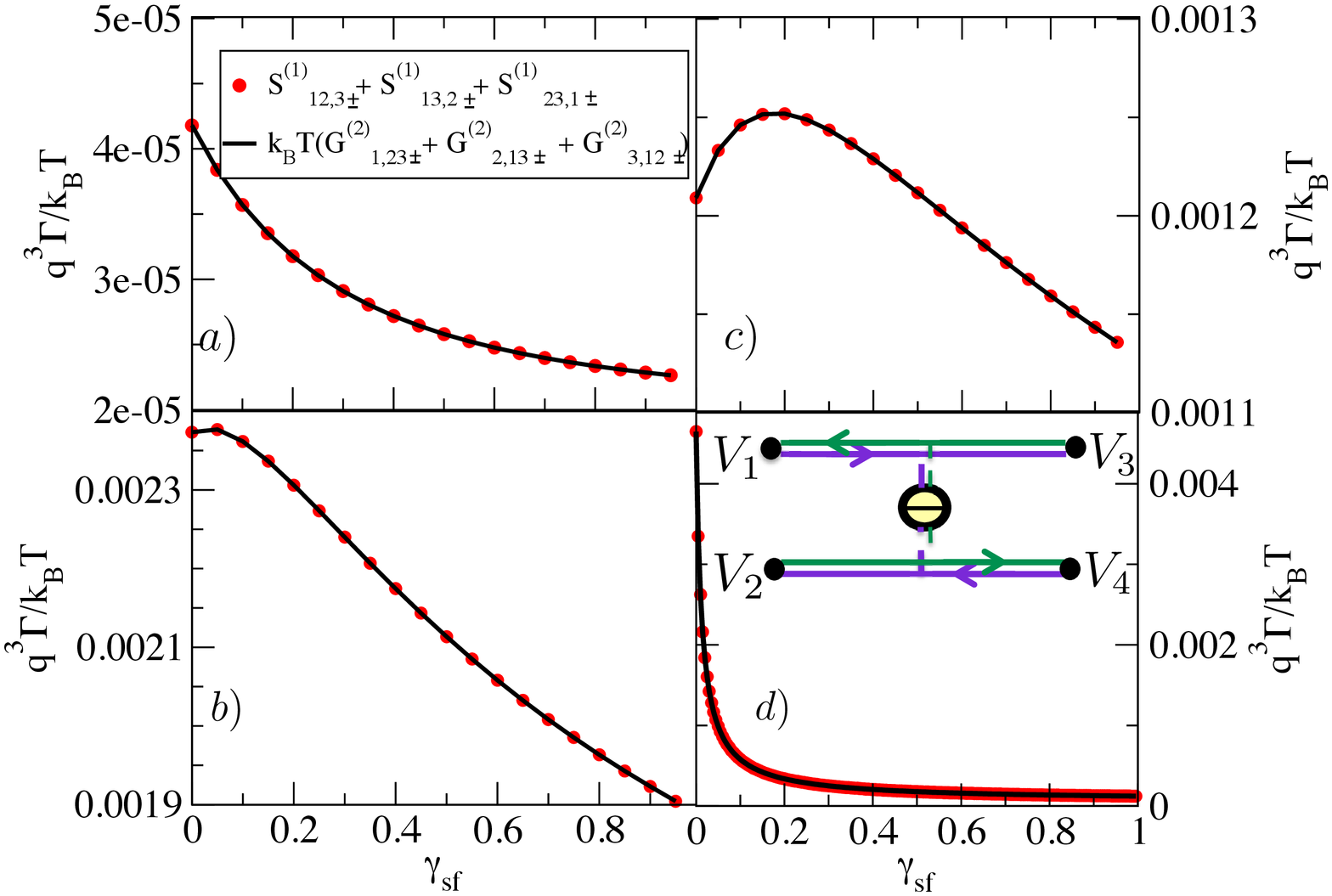}
\caption[Verification of spintronic fluctuation relations]
{Verification of spintronic fluctuation relations as a function of $\gamma_{sf}$ in the presence of magnetic interactions, $B$, and $p$ for different values of polarization $p$:  (a) $p=0$, (b) $p=0.25$, (c) $p= 0.5$, and (d) $p=0.75$. Upper helical modes: left (right) movers
are spin up (down) injected with voltage $V_{1(3)}$.  Lower helical modes: right (left) movers 
are spin up (down) injected with voltage $V_{2(4)}$. Parameters: $\Gamma = 1$, $q^2/[4(C_1+C_2)]= 40\Gamma$, $\veps_d = 0$,  $k_BT = 5\Gamma$, $g\mu_B B=0.1\Gamma$
and capacitance asymmetry $\eta=0.5$. Note that in our chiral system spin indices
are included in the lead indices for the fluctuation relations.}
\end{figure}
\paragraph{Acknowledgments.}
Work supported by MINECO Grants Nos.\ FIS2011-23526 and CSD2007-00042 (CPAN). We thank M. Esposito for fruitul discussions about the general role of the local detailed balance condition in FTs. We also thank M. B\"{u}ttiker and R. S\'{a}nchez for carefully reading of the manuscript and their suggestions and comments.


\begin{thebibliography}{99}

\bibitem{PhysRevLett.78.2690}
\bibinfo{author}{{C.}~{Jarzynski}},
  \bibinfo{journal}{Phys. Rev. Lett.} \textbf{\bibinfo{volume}{78}},
  \bibinfo{pages}{2690} (\bibinfo{year}{1997}).
  
\bibitem{PhysRevE.60.2721}
\bibinfo{author}{{G.~E.} {Crooks}},
  \bibinfo{journal}{Phys. Rev. E} \textbf{\bibinfo{volume}{60}},
  \bibinfo{pages}{2721} (\bibinfo{year}{1999}).

\bibitem{RevModPhys.81.1665}
\bibinfo{author}{{M.}~{Esposito}},
  \bibinfo{author}{{U.}~{Harbola}}, {and}
  \bibinfo{author}{{S.}~{Mukamel}},
  \bibinfo{journal}{Rev. Mod. Phys.} \textbf{\bibinfo{volume}{81}},
  \bibinfo{pages}{1665} (\bibinfo{year}{2009}).

\bibitem{levitov}
L. S. Levitov and G. B. Lesovik, Pis’ma Zh. Eksp. Teor.
Fiz. {\bf 58}, 225 (1993).

\bibitem{PhysRevLett.88.196801}
\bibinfo{author}{{Y.~V.} {Nazarov}} {and}
  \bibinfo{author}{{D.~A.} {Bagrets}},
  \bibinfo{journal}{Phys. Rev. Lett.} \textbf{\bibinfo{volume}{88}},
  \bibinfo{pages}{196801} (\bibinfo{year}{2002}).

\bibitem{PhysRevLett.90.206801}
\bibinfo{author}{{S.}~{Pilgram}},
  \bibinfo{author}{{A.~N.} {Jordan}},
  \bibinfo{author}{{E.~V.} {Sukhorukov}},
  {and}
  \bibinfo{author}{{M.}~{B\"uttiker}},
  \bibinfo{journal}{Phys. Rev. Lett.} \textbf{\bibinfo{volume}{90}},
  \bibinfo{pages}{206801} (\bibinfo{year}{2003}).
  
  \bibitem{1742-5468-2006-01-P01011}
\bibinfo{author}{{D.}~{Andrieux}} {and}
  \bibinfo{author}{{P.}~{Gaspard}},
  \bibinfo{journal}{J. Stat. Mech.}
  \bibinfo{volume}{P01011} (\bibinfo{year}{2006}).
  
  \bibitem{PhysRev.32.110}
\bibinfo{author}{{H.}~{Nyquist}},
  \bibinfo{journal}{Phys. Rev.} \textbf{\bibinfo{volume}{32}},
  \bibinfo{pages}{110} (\bibinfo{year}{1928}).
  
  \bibitem{PhysRev.32.97}
\bibinfo{author}{{J.~B.} {Johnson}},
  \bibinfo{journal}{Phys. Rev.} \textbf{\bibinfo{volume}{32}},
  \bibinfo{pages}{97} (\bibinfo{year}{1928}).

\bibitem{JPSJ.12.570}
\bibinfo{author}{{R.}~{Kubo}},
  \bibinfo{journal}{J. Phys. Soc. Jpn.}
  \textbf{\bibinfo{volume}{12}}, \bibinfo{pages}{570} (\bibinfo{year}{1957}).

\bibitem{Einstein.1905}
\bibinfo{author}{{A.}~{Einstein}},
  \bibinfo{journal}{Ann. Phys. Lpz.} \textbf{\bibinfo{volume}{322}},
  \bibinfo{pages}{549}
  (\bibinfo{year}{1905}).

\bibitem{PhysRevB.72.235328}
\bibinfo{author}{{J.}~{Tobiska}} {and}
  \bibinfo{author}{{Y.~V.} {Nazarov}},
  \bibinfo{journal}{Phys. Rev. B} \textbf{\bibinfo{volume}{72}},
  \bibinfo{pages}{235328} (\bibinfo{year}{2005}).

\bibitem{PhysRevLett.101.046802}
\bibinfo{author}{{R.~D.} {Astumian}},
  \bibinfo{journal}{Phys. Rev. Lett.} \textbf{\bibinfo{volume}{101}},
  \bibinfo{pages}{046802} (\bibinfo{year}{2008}).

\bibitem{PhysRevB.78.115429}
\bibinfo{author}{{K.}~{Saito}} {and}
  \bibinfo{author}{{Y.}~{Utsumi}},
  \bibinfo{journal}{Phys. Rev. B} \textbf{\bibinfo{volume}{78}},
  \bibinfo{pages}{115429} (\bibinfo{year}{2008});
  Phys. Rev. B \textbf{79}, 235311 (2009);
  J. Phys. Conf. Series \textbf{200}, 052030 (2010).

\bibitem{PhysRevLett.101.136805}
\bibinfo{author}{{H.}~{F\"orster}} {and}
  \bibinfo{author}{{M.}~{B\"uttiker}},
  \bibinfo{journal}{Phys. Rev. Lett.} \textbf{\bibinfo{volume}{101}},
  \bibinfo{pages}{136805} (\bibinfo{year}{2008});
AIP Conf. Proc. \textbf{1129}, 443 (2009).

\bibitem{san09}
\bibinfo{author}{{D.}~{S\'anchez}},
  \bibinfo{journal}{Phys. Rev. B} \textbf{\bibinfo{volume}{79}},
  \bibinfo{pages}{045305} (\bibinfo{year}{2009}).
  
  \bibitem{san10}
R. S\'anchez, R. L\'opez, D. S\'anchez, and M. B\"uttiker,
Phys. Rev. Lett. \textbf{104}, 076801 (2010).

\bibitem{lim10}
J.S. Lim, D. S\'anchez, and R. L\'opez,
Phys. Rev. B \textbf{81}, 155323 (2010);
AIP Conf. Proc. \textbf{1129}, 435 (2009).
  
\bibitem{gol11}
D. S. Golubev, Y. Utsumi, M. Marthaler, and G. Sch\"on,
Phys. Rev. B \textbf{84}, 075323 (2011).

\bibitem{bul11}
G. Bulnes Cuetara, M. Esposito, and P. Gaspard,
Phys. Rev. B \textbf{84}, 165114 (2011).

\bibitem{kra11}
T. Krause, G. Schaller, and T. Brandes,
Phys. Rev. B {\bf 84}, 195113 (2011). 

\bibitem{gan11}
  S. Ganeshan and N. A. Sinitsyn,
Phys. Rev. B \textbf{84}, 245405 (2011).

\bibitem{fab07}
J. Fabian, A. Matos-Abiague, C. Ertler, P. Stano, and I. Zutic, 
Acta Physica Slovaca \textbf{57}, 565 (2007).


  \bibitem{PhysRevLett.104.080602}
\bibinfo{author}{{S.}~{Nakamura}},
  \bibinfo{author}{{Y.}~{Yamauchi}},
  \bibinfo{author}{{M.}~{Hashisaka}},
  \bibinfo{author}{{K.}~{Chida}},
  \bibinfo{author}{{K.}~{Kobayashi}},
  \bibinfo{author}{{T.}~{Ono}},
  \bibinfo{author}{{R.}~{Leturcq}},
  \bibinfo{author}{{K.}~{Ensslin}},
  \bibinfo{author}{{K.}~{Saito}},
  \bibinfo{author}{{Y.}~{Utsumi}},
  \bibinfo{author}{{A.C.}~{Gossard}},
\bibinfo{journal}{Phys. Rev. Lett.}
  \textbf{\bibinfo{volume}{104}}, 080602 (\bibinfo{year}{2010});
  Phys. Rev. B \textbf{83}, 155431 (2011).


\bibitem[{{S\'anchez and
  B\"uttiker}(2004)}]{PhysRevLett.93.106802}
\bibinfo{author}{{D.}~{S\'anchez}} {and}
  \bibinfo{author}{{M.}~{B\"uttiker}},
  \bibinfo{journal}{Phys. Rev. Lett.} \textbf{\bibinfo{volume}{93}},
  \bibinfo{pages}{106802} (\bibinfo{year}{2004});
  Phys. Rev. B {\bf 72}, 201308(R) (2005).
  
 \bibitem{ada06} 
I. Adagideli, G. E. W. Bauer, and B. I. Halperin,
Phys. Rev. Lett. \textbf{97}, 256601 (2006).

\bibitem[{{Hasan and Kane}(2010)}]{RevModPhys.82.3045}
\bibinfo{author}{{M.~Z.} {Hasan}} {and}
  \bibinfo{author}{{C.~L.} {Kane}},
  \bibinfo{journal}{Rev. Mod. Phys.} \textbf{\bibinfo{volume}{82}},
  \bibinfo{pages}{3045} (\bibinfo{year}{2010}).
  
  
\bibitem[{{Kloeffel et~al.}(2011){Kloeffel, Trif, and
  Loss}}]{PhysRevB.84.195314}
\bibinfo{author}{{C.}~{Kloeffel}},
  \bibinfo{author}{{M.}~{Trif}}, {and}
  \bibinfo{author}{{D.}~{Loss}},
  \bibinfo{journal}{Phys. Rev. B} \textbf{\bibinfo{volume}{84}},
  \bibinfo{pages}{195314} (\bibinfo{year}{2011}).


\bibitem[{{Klinovaja et~al.}(2011){Klinovaja,
  Schmidt, Braunecker, and Loss}}]{PhysRevLett.106.156809}
\bibinfo{author}{{J.}~{Klinovaja}},
  \bibinfo{author}{{M.~J.} {Schmidt}},
  \bibinfo{author}{{B.}~{Braunecker}},
  {and} \bibinfo{author}{{D.}~{Loss}},
  \bibinfo{journal}{Phys. Rev. Lett.} \textbf{\bibinfo{volume}{106}},
  \bibinfo{pages}{156809} (\bibinfo{year}{2011}).
  
\bibitem[{{Tsukagoshi et~al.}(2011){Tsukagoshi,
  Alphenaar, and Ago}}]{ferroleads}
\bibinfo{author}{{K.}~{Tsukagoshi}},
  \bibinfo{author}{{B.~W.} {Alphenaar}},
  {and} \bibinfo{author}{{H.}~{Ago}},
  \bibinfo{journal}{Nature} \textbf{\bibinfo{volume}{401}},
  \bibinfo{pages}{572} (\bibinfo{year}{1999}).

\bibitem[{{Carsten}(2011)}]{spinhallQD}
\bibinfo{author}{{C.}~{Timm}},
  \bibinfo{journal}{arXiv:1111.2245}  (\bibinfo{year}{2011}).
  
  	\bibitem{vai2007}
	M.H. Vainstein and J.M. Rub\'{\i}, Phys. Rev. E {\bf 75}, 031106 (2007).
	
	\bibitem{wang2011}
	C. Wang and D. E. Feldman, Phys. Rev. B 84, 235315 (2011).

\bibitem[{{Christen and B\"uttiker}(1996)}]{PhysRevB.53.2064}
\bibinfo{author}{{T.}~{Christen}} {and}
  \bibinfo{author}{{M.}~{B\"uttiker}},
  \bibinfo{journal}{Phys. Rev. B} \textbf{\bibinfo{volume}{53}},
  \bibinfo{pages}{2064} (\bibinfo{year}{1996}).
  
\bibitem{mac99}
A.H. MacDonald,
Phys. Rev. Lett. \textbf{83}, 3262 (1999).

\end{thebibliography}
\end{document}